\documentclass[superscriptaddress,onecolumn,showpacs,preprintnumbers,amsmath,amssymb]{revtex4}
\usepackage{epsfig}
\usepackage{dcolumn}

\usepackage[latin1]{inputenc}
\usepackage{amsmath}
\usepackage{psfrag}

\usepackage{color}
\newcommand{\chem}[1]{\ensuremath{\mathrm{#1}}}
\newcommand{\un}[1]{\ensuremath{\unskip\,\mathrm{#1}}}

\begin{document}
\title{Noisy saltatory spike propagation: the breakdown of signal
  transmission due to channel noise}
\author{Yunyun Li}
\affiliation{Institut f\"ur Physik, Universit\"at Augsburg,
  Universit\"atsstr. 1, 86159 Augsburg, Germany}

\author{Gerhard Schmid}
\affiliation{Institut f\"ur Physik, Universit\"at Augsburg,
  Universit\"atsstr. 1, 86159 Augsburg, Germany}

\author{Peter H\"anggi}
\affiliation{Institut f\"ur Physik, Universit\"at Augsburg,
  Universit\"atsstr. 1, 86159 Augsburg, Germany}

\date{\today}

\begin{abstract}
Noisy saltatory spike propagation along myelinated axons is studied within a
stochastic Hodgkin-Huxley model. The {\it intrinsic} noise
(whose strength is inverse proportional to the nodal membrane size) arising from fluctuations
of the number of open ion channels  influences the dynamics of the
membrane potential in a node of Ranvier where the sodium ion channels
are predominantly localized. The nodes of Ranvier are linearly coupled. As the
measure for the signal propagation reliability we focus on the ratio
between the number of initiated spikes and the transmitted spikes. This
work supplements our earlier study [A. Ochab-Marcinek, G. Schmid,
I. Goychuk and P. H\"anggi, Phys. Rev E {\bfseries 79}, 011904 (2009)]
towards stronger channel noise intensity and supra-threshold
coupling. 
For strong supra-threshold coupling the transmission reliability
decreases with increasing channel noise level until the causal
relationship is completely lost and a breakdown of the spike
propagation due to the intrinsic noise is observed.    
\end{abstract}
\maketitle
\section{Introduction}
\label{sec:intro}

As the fast propagation of action potentials along the axon is
of fundamental importance in the nervous system, e.g. for the successful
evolution to large body sizes of organisms, or the information
processing in the brain, the study of the propagation mechanisms is of
great interest for neuroscientists, physiologists and physicists
~\cite{bb}. Since the empirical modeling proposed by Hodgkin and Huxley in
1952 ~\cite{Hodgkin1952}, the neuronal firing dynamics with respect to
spike generation and signal propagation  are successfully studied within
this deterministic modeling. In recent days,  issues relating to the constructive role of
noise  on these dynamics were addressed \cite{white2000,lindner}. The
noise-assisted enhancement in weak signals transmission, transduction
or detection has been investigated, e.g. in the context of noise
supported wave propagation in sub-excitable media~\cite{sagues} or in
excitable systems~\cite{lindner}. The conductance
fluctuation of the neuronal membranes which arises from random channel opening and closing
can, {\itshape a priori} not be neglected~\cite{Lecar}. Indeed, it was
shown, that this intrinsic {\itshape channel noise} \cite{white2000} can lead to generation of
so-called {\itshape spontaneous action potentials}, to
synchronization phenomena like {\itshape stochastic
  resonance}~\cite{Gammaitoni,Anishenkov,hanggicpc,SchmidEPL,JungEPL}
and {\itshape coherence resonance}~\cite{schmid2004,schmid2006}
and to synchronization of
ion channel clusters~\cite{Zeng}.

Even the saltatory spike propagation which results from a
highly non-uniform distribution of the ion channel, can take benefit
of the intrinsic channel noise as we have shown recently in Ref.~\cite{Anna}. The saltatory
spike propagation occurs in myelinated axons where the activating
sodium ion channels are concentrated at the nodes of Ranvier, which
are separated by segments sheathed with myelin.  This results in a
much faster propagation speed in myelinated axons than that in un-myelinated axons with
constant ion channel density~\cite{Koch,Keener}.

With this work, we extend our prior study~\cite{Anna} on  the effect of
channel noise on the propagation of action potentials along myelinated
axons. 
In terms of transmission reliability we are discussing the influence of the
coupling strength between neighboring nodes of Ranvier and that of
strong intrinsic noise.

\section{Model}
\label{sec:model}

In order to model the signal transmission along  myelinated axons, we
consider a compartmental stochastic Hodgkin-Huxley model \cite{Anna}.
Accordingly, each node of Ranvier is modeled by a stochastic
generalization of the Hodgkin-Huxley model, which extends the
applicability of the original
Hodgkin-Huxley model \cite{Hodgkin1952} towards stochastic dynamics of the
membrane potential of finite-size ion channel clusters
\cite{white2000,Fox1994}. Each node of Ranvier couples linearly to its
nearest neighbors. Consequently, the membrane dynamics $V_{i}$ at the
$i$th node of Ranvier reads (with $i = 0,1,2,..., N-1$, where $N$
corresponds to the total number of axonal nodes of Ranvier):

\begin{figure}
\includegraphics[width=\linewidth]{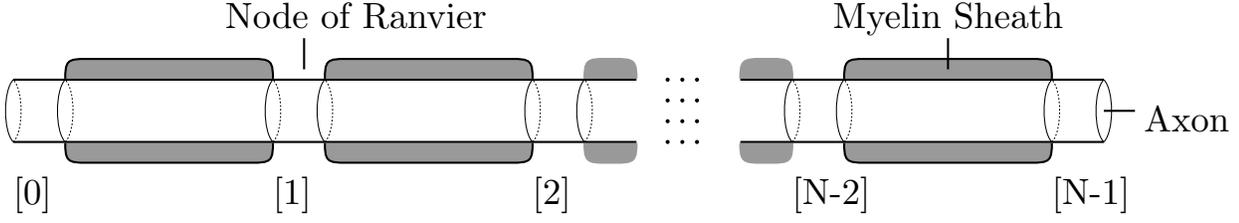}
\caption{Sketch of the myelinated axon: Each node of Ranvier is
  treated within a stochastic generalization of the Hodgkin-Huxley
  model and is bi-linearly coupled to the nearest neighboring nodes of
  Ranvier.}
\label{fig:1}       
\end{figure}

\begin{subequations}
  \label{eq:compartmentmodel}
\begin{align}
  \label{eq:comp_mod}
  C \frac{\mathrm d}{\mathrm{d} t} V_{i} &= I_{i,
    \mathrm{ionic}}(V_{i})  + I_{i, \mathrm{inter}}(t) + I_{i, \mathrm{ext}}(t)\, , \quad \text{for
  } i=0,1,2,...,N-1\, , \\
\intertext{ with the ionic membrane current (per unit area) within the
  $i$th node of Ranvier given by the Hodgkin-Huxley model \cite{Hodgkin1952}}
  \label{eq:voltage-equation}
  I_{i,\mathrm{ionic}}(V_{i}) &= -G_{\chem{K}}(n_{i})\ (V_{i}-E_{\chem{K}})
 -G_{\chem{Na}}(m_{i},h_{i})\ ( V_{i} -  E_{\chem{Na}})
  -G_{\chem{L}}(V_{i} -  E_{\chem{L}})\, ,\\
\intertext{ the inter-nodal currents}
  \label{eq:internodal-currents}
    I_{i, \mathrm{inter}}(t)&=
    \begin{cases}
  \kappa \left( V_{i+1} - V_{i}\right) & \text{for }i=0\, ,\\
  \kappa \left( V_{i-1} - V_{i}\right) & \text{for }i=N-1\, ,\\
\kappa
  \left( V_{i-1}-2 V_{i} + V_{i+1}\right) & \text{elsewhere\, }
    \end{cases}
\end{align}
\end{subequations}
and the external current stimuli $I_{i,\mathrm{ext}}(t)$
  at the $i$th node of Ranvier. In Eq.~(\ref{eq:comp_mod}), $C$ denotes
  the capacity of the axonal membrane per unit area and is given by
  $C=1 \un{\mu F/cm}^{2}$. The coupling strength
between next-neighboring nodes is characterized by $\kappa$,
cf. Eq.~(\ref{eq:internodal-currents}).  According to Hodgkin-Huxley
model~\cite{Hodgkin1952}, the reversal potentials read for the sodium current: $E_{\chem{Na}}
= 50\un{mV}$, for the potassium current: $E_{\chem{K}} = -77\un{mV}$
and for the leakage: $E_{L} = -54.4 \un{mV}$. The conductances per
unit area are given by:
\begin{equation}
  \label{eq:conductances-hodgkinhuxley}
  G_{\chem{K}}(n_{i})=g_{\chem{K}}^{\mathrm{max}}\  n_{i}^{4} , \quad
  G_{\chem{Na}}(m_{i},h_{i})=g_{\chem{Na}}^{\mathrm{max}}\ m_{i}^{3} h_{i}\, .
\end{equation}
and the constant leakage conductance $G_{\chem{L}}=0.3\un{mS/cm^{2}}$. In
Eq.~(\ref{eq:conductances-hodgkinhuxley}), $g_{\chem K}^{\mathrm{max}}$ and
$g_{\chem{Na}}^{\mathrm{max}}$ denote the maximum potassium and
sodium conductances per unit area, when all ion channels within the corresponding
node are open. For simplicity, we assume that every axonal node has
the same kinetics, i.e. the same number of sodium and potassium ion
channels. So the maximum potassium and sodium conductances $g_{\chem
  {K}}^{\mathrm{max}}= 36 \un{mS/cm^{2}}$ and
$g_{\chem {Na}}^{\mathrm{max}}= 120 \un{mS/cm^{2}}$ are identical
constants for every node of Ranvier.

The gating variables $n_{i},\ m_{i}$ and $h_{i}$ in Eqs.~\eqref{eq:voltage-equation} and
\eqref{eq:conductances-hodgkinhuxley}, describe the open probabilities
of the ion channel gates in the $i$th node, and undergo a stochastic
process which stems from a birth-and-death-like process of the gating
dynamics.  The dynamics of the gating variables are voltage dependent,
and are governed by the set of (Ito)-stochastic differential equations \cite{Fox1994,Tuckwell1987,HanggiThomas_a},

\begin{align}
  \label{eq:stochasticgates}
  \frac{\mathrm{d}}{\mathrm{dt}}x_i =
  \alpha_{x}(V_{i})\ (1-x_i)-\beta_{x}(V_{i})\ x_i + \xi_{i, x}(t)\, ,
\end{align}

with $x=m,h,n$. Here, $\xi_{i, x}(t)$ are Gaussian white
noise with vanishing mean and vanishing cross-correlations.
For the same node with the nodal membrane size ${\cal A}$ the
non-vanishing noise correlations take the following form:

\begin{subequations}
\label{eq:noisecor}
\begin{align}
  \label{eq:correlator-a}
  \langle \xi_{i, m}(t)\, \xi_{i, m}(t') \rangle &=
  \frac{1}{{\cal A} \rho_{\chem{Na}}}\
   [\alpha_{m}(V_{i})\, (1-m_i) + \beta_{m}(V_{i})\, m_i]\ \delta(t -t')\, ,  \\
  \label{eq:correlator-b}
  \langle \xi_{i, h}(t)\, \xi_{i, h}(t') \rangle &=  \frac{1}{{\cal A} \rho_{\chem{Na}}}\
  [ \alpha_{h}(V_{i})\, (1-h_i) + \beta_{h}(V_{i})\, h_i]\ \delta(t -t')\, , \\
  \label{eq:correlator-c}
  \langle \xi_{i, n}(t)\, \xi_{i, n}(t') \rangle &=  \frac{1}{{\cal A} \rho_{\chem{K}}}\
  [ \alpha_{n}(V_{i})\, (1-n_i)+
    \beta_{n}(V_{i})\, n_i ]\ \delta(t -t')  \, ,
\end{align}
\end{subequations}
where the ion channel
densities are $\rho_{\chem{Na}}= 60 \un{\mu m^{-2}}$ and
$\rho_{\chem{K}} = 18 \un{\mu m^{-2}}$. The noise strength
is decided by the nodal membrane size ${\cal A}$ which is the
same for all nodes. In Eq.~\ref{eq:stochasticgates}, the dynamics of the opening and closing rates $\alpha_{x}(V)$ and
$\beta_{x}(V)\; (x=m,h,n)$ are taken at  $T=6.3 \un{^{\circ} C}$. They
depend on the local membrane potential $V$ and read (with numbers
given in units of $[\text{m} V]$)~\cite{Hodgkin1952,goychukpnas}:
\begin{subequations}
\label{eq:gating}
\begin{align}
  \label{eq:rates-m}
  \alpha_{m}(V) &=  \frac{0.1(V+40)}{1 - \exp\left\{-\, (V+40)/10
    \right\}}\, , \\
  \beta_{m}(V)  &= 4\, \exp\left\{ -\, (V+65)/18 \right\}\, , \\
  \alpha_{h}(V) &= 0.07  \, \exp\left\{ -\, (V+65)/20 \right\}\, , \\
  \beta_{h}(V)  &= \frac{1}{1 + \exp\left\{-\, (V+35)/10 \right\} }\, ,  \\
  \alpha_{n}(V) &= \frac{0.01(V + 55)}{1 - \exp \left\{-\, (V + 55)
      / 10 \right\}}\, ,\\
  \label{eq:rates-n}
  \beta_{n}(V)  &= 0.125 \, \exp\left\{ -\, (V+65)/80 \right\}\, .
\end{align}
\end{subequations}

\section{Spike transmission}
\label{sec:results}

In order to analyze the transmission reliability we exemplary consider
a chain consisting of ten nodes of Ranvier, i.e. $N=10$ and numerically
simulate Eq.~\eqref{eq:compartmentmodel}. By applying a constant
current stimulus on the first node only, i.e. we set $I_{0,\mathrm{ext}} = 12
\un{\mu A/cm^{2}}$ and $I_{i,\mathrm{ext}} = 0 $ for $i=1,...,N-1$, we find that
action potentials are periodically produced that propagate along the
transmission line. We define the transmission reliability coefficient $\cal R$ in
steady state by the ratio
of the number of action potentials arriving at the terminal node
${\cal N}_{9}$ to
those generated in the initial node ${ \cal N}_{0}$:
\begin{align}
  \label{eq:ratio}
  \text{transmission reliability coefficient: }{\cal R}  =
  \frac{{\cal N}_{9}}{{\cal N}_{0}}
\end{align}
 The occurrence of a spike in the membrane potential $V_{i}(t)$ is identified
by an upward-crossing of the detection barrier at $0.0 \un{mV}$ (
Further
details on the numerics can be found in Ref.~\cite{Anna}).
The spike occurrences $t_{i}^{j}$ with $j=1,...,{\cal N}_{i}$ where ${ \cal N}_{i}$ indicates the number of spikes appeared on the specific node $i$ define point processes $u_{i}(t)=\sum_{j=1}^{{\cal
    N}_{i}} \, \delta ( t - t_{i}^{j})$.

\subsection{Deterministic dynamics}
\label{sec:detdyn}

\begin{figure}
  \centering
  \includegraphics{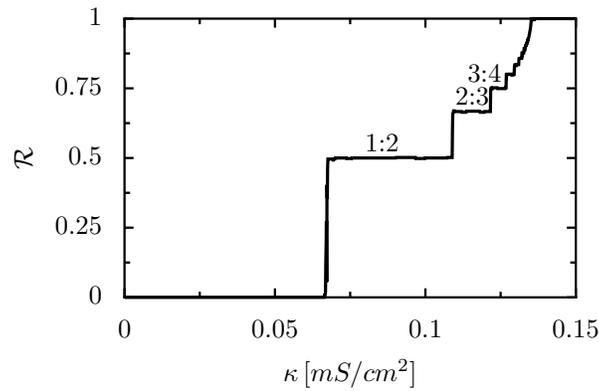}
  \caption{Deterministic transmission reliability: The dependence of
    the transmission reliability ${\cal R}$ on the inter-nodal coupling strength
    $\kappa$ is depicted for the deterministic case. For sub-threshold
    $\kappa \lesssim 0.067 \un{mS/cm^{2}}$ no spike propagation
    is observed, i.e. ${\cal R}=0$. Perfect spike transmission,
    i.e. ${\cal R}=1$, is found for the
    supra-threshold case $\kappa \gtrsim 0.136 \un{mS/cm^{2}}$. In the intermediate range rational numbers $k:l$ for
    the transmission reliability are found. }
  \label{fig:detdyn}
\end{figure}

We start with the deterministic limit which is formally achieved in
the limit $\mathcal A \to \infty$, using $N=10$. In this case, the
transmission reliability ${\cal R}$ depends solely on the inter-nodal
coupling strength $\kappa$ and exhibits distinguished transmission
patterns \cite{Anna}.

The dependence of the transmission reliability ${\cal R}$ on the coupling
strength is depicted in Fig.\ref{fig:detdyn}. For sub-threshold
coupling, i.e. $\kappa \lesssim 0.067 \un{mS/cm^{2}}$
the ratio equals zero and no spike propagation to the final node is observed.
Contrary, for sufficiently large coupling parameter, i.e. $\kappa
\gtrsim 0.136 \un{mS/cm^{2}}$, each generated
action potential propagates along the axon and arrives at the final
node, i.e. ${\cal R} = 1$. Discrete, rational transmission
patterns $k:l$ appear for intermediate values of the coupling
parameter.

\subsection{Channel noise effects}

When considering finite sizes ${\cal A}$ of the nodes of Ranvier, the
staircase-like dependence of the transmission reliability parameter ${\cal
  R}$, which is  depicted in Fig.~\ref{fig:detdyn} for the
deterministic case (i.e. for the case of an infinite size of the nodes of
Ranvier), turns in a continuous dependence (not shown).

\begin{figure}[t]
  \centering
  \includegraphics{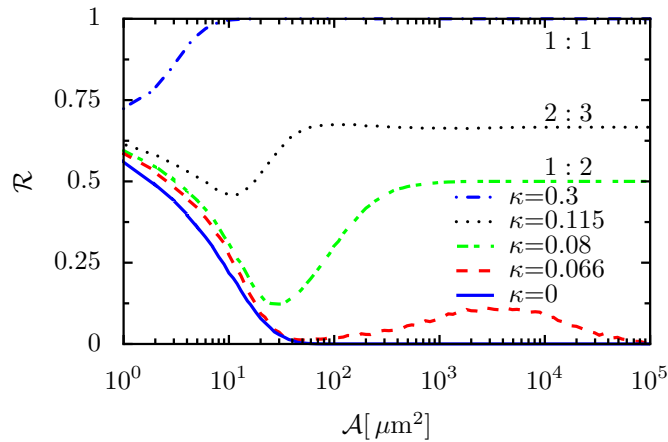}
  \caption{(color online) Transmission reliability in presence of channel noise: The
    transmission reliability coefficient ${\cal R}$ is plotted against
  the membranal size of the node of Ranvier for different strengths of
  the inter-nodal coupling (the coupling strengths are given in units of
  \un{mS/cm^{2}}) . The
  data confirms the result of 
  noise-assisted spike propagation for sub-threshold coupling (see
  dashed red line) \cite{Anna}. Moreover, the minimum in ${\cal R}$
  indicates the cross-over from a
  stochastic, uncorrelated spiking of the initial and final nodes towards a causal, noisy
  spike transmission from the initial to the final node. }
  \label{fig:stochdyn}
\end{figure}

\begin{figure}
\centering
\epsfig{figure=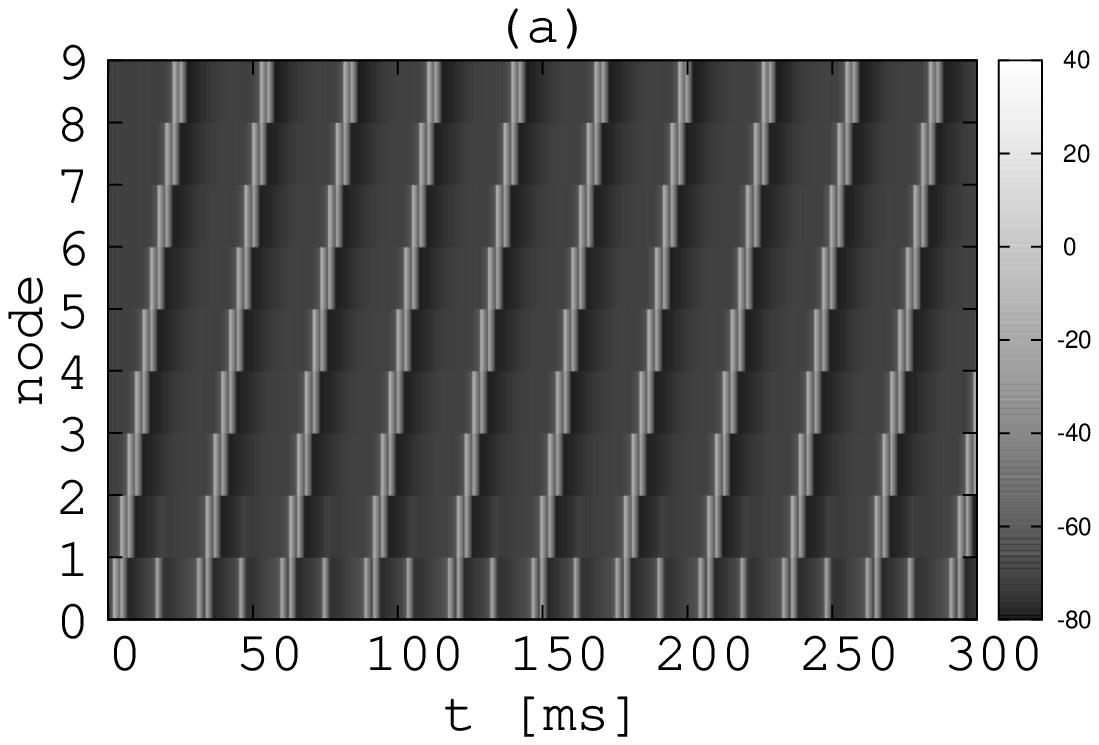, scale=0.35}
\epsfig{figure=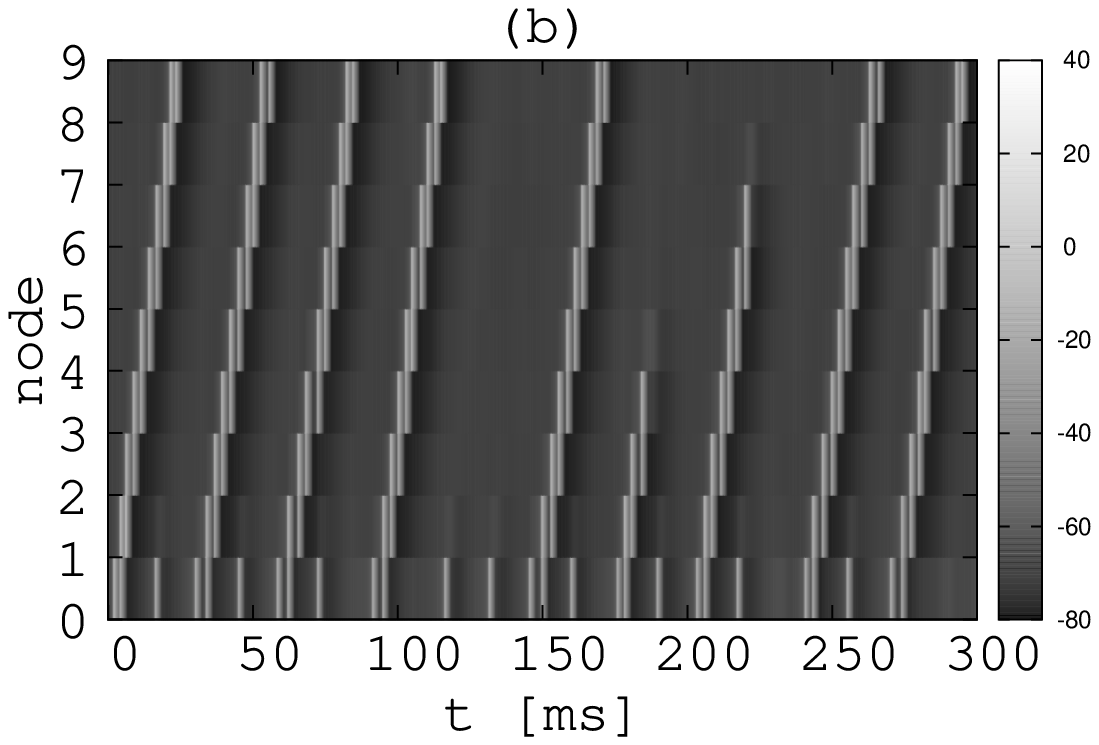, scale=0.35}
\epsfig{figure=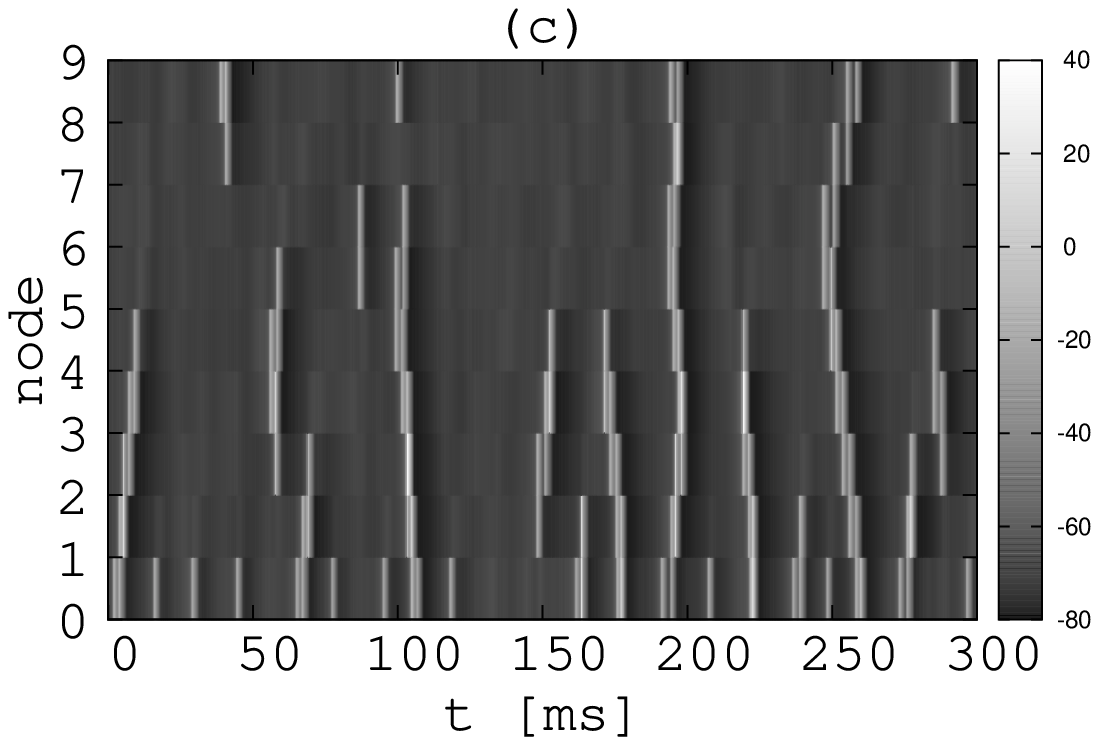, scale=0.35}
\caption{Spatial-temporal transmission patterns of spike propagation:
  the spatial-temporal evolution of the membrane dynamics at
  different axonal nodes of Ranvier is plotted for an intermediate
  coupling strength $\kappa = 0.08 \un{mS/cm^{2}}$ and various nodal
  membrane sizes: (a) ${\cal A} = 3\cdot 10^{4} \un{\mu m^{2}}$ (low
  channel noise level); (b) ${\cal A} = 100 \un{\mu m^{2}}$ (intermediate
  channel noise level) and (c) ${\cal A} = 10 \un{\mu m^{2}}$ (high
  channel noise level). Next to each panel, there is a color bar
  indicating the actual value of the membrane potential.
  The action potentials are created by a
  current stimulus at the initial node ($"0"$). In case of low noise
  level, cf. panel (a), the deterministic
  transmission pattern of $2:1$ is clearly visible. In contrast, in
  the strong noise limit, due to the dominating spontaneous spiking,
  irregular transmission is found.  }
\label{fig:vt}
\end{figure}

In the limit of strong intrinsic noise, i.e. for the finite size of the nodes of Ranvier
approaching formally zero, the channel noise reigns the dynamics of the membrane
potential of each node and the influence of the bi-linear coupling
becomes negligible. Consequently, in this limit 
the causal relationship between spiking in the first and that one of
the terminal node gets lost. The number of spike
occurrences at the final node does not markedly depend on the actual value of the
coupling parameter and the ratio ${\cal R}$ tends to a value independent of coupling strength
 $\kappa$, cf. Fig.~\ref{fig:stochdyn}. Note, that in this case,
the information transfer from an initial to a final node fades towards
zero as the correlation between spikes in the initial and the final node
diminishes. Moreover, the spontaneous spikes stimulates via the coupling the neighboring nodes
and consequently a transmission to both sides occurs, cf. Fig.~\ref{fig:vt} (c).
In the limit $\kappa \to 0$, the initial and the terminal node are
spiking independently. Consequently, ${\cal R}$ is given by the ratio
of the number of spontaneous spikes occurring within the non-stimulated, stochastic Hodgkin-Huxley
dynamics and the  number of spikes occurring within the stochastic
Hodgkin-Huxley dynamics complemented by a constant current stimulus of
$I=12 \un{\mu A/cm^{2}}$. Surely, the causal relationship is lost in
this case. This becomes evident when considering the cross-correlation
between the spiking of the initial and that of the final node (see below).  

With increasing size of the nodes of the Ranvier $\cal A$, i.e. with
decreasing channel noise strength, the transmission
becomes more regular, cf. Fig.~\ref{fig:vt} (a) and
(b). The causal relationship is restored in this limit and the
transmission coefficient ${\cal R}$ tends to its deterministic limit.
This cross-over from the stochastic
non-causal firing regime to the regime of noisy spike transmission
depends on the coupling strength. For smaller coupling strengths, this
cross-over occurs at larger nodal membrane sizes  ${\cal A}$,
cf. Fig.~\ref{fig:stochdyn}.

Starting out from the weak noise limit, an increase in the noise
strength results in a noise-assisted spike propagation phenomenon for 
sub-threshold coupling as pointed out before with
Ref.~\cite{Anna}. The transmission reliability 
exhibits a maximum, indicating an optimal, noise assisted spike propagation,
cf. Fig.~\ref{fig:stochdyn}. In this case the coupling between the
nodes does not result in an efficient propagation of the spikes and the presence of intrinsic
noise is necessary to overcome the threshold for excitation.
However, for the supra-threshold coupling, the channel noise leads to
noise-induced propagation failures and the transmission reliability
coefficient ${\cal R}$ firstly decreases with increasing noise
level, cf. Fig.~\ref{fig:stochdyn}. The observed increase of ${\cal R}$ for
strong intrinsic noise is attributed to the cross-over to the
stochastic, non-causal firing regime accompanied by uncorrelated
spiking in the initial and final nodes. 

\begin{figure}
\centering
\epsfig{figure=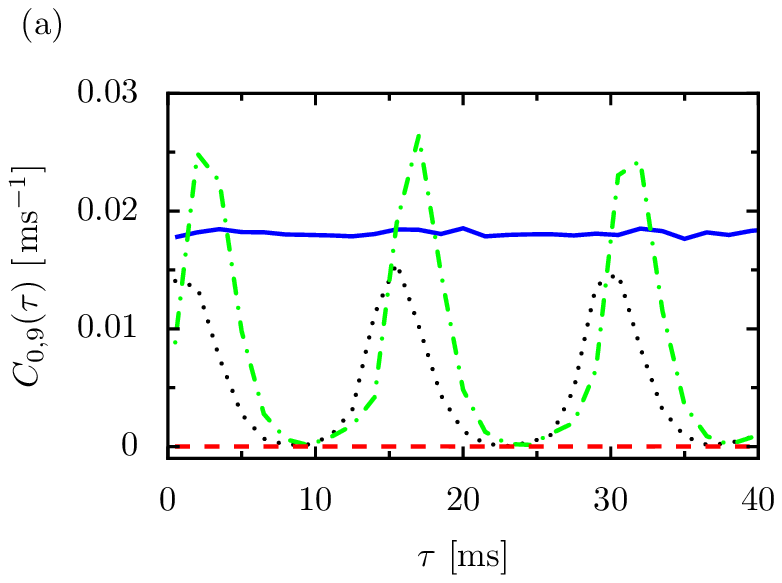, scale=0.8}
\epsfig{figure=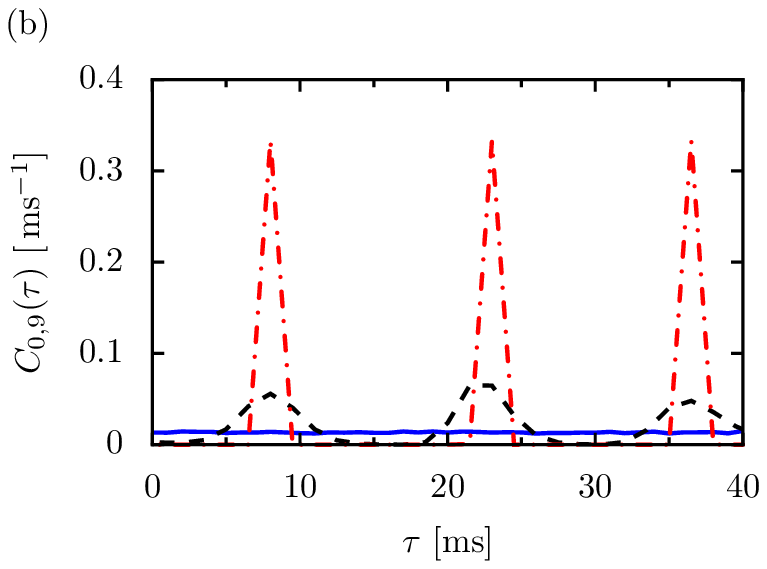, scale=0.8}
\caption{(Color online) Initial-final-node spiking correlation: The spike
  correlation $C_{0,9} (\tau)$ between the spiking at the initial node and that of the
  final node, cf. Eq.~(\ref{eq:prop}), is plotted for different coupling strengths $\kappa$ and
  membrane sizes ${\cal A}$: In panel (a) for sub-threshold coupling $\kappa =
  0.066 \un{mS/cm^{2}}$ and membrane sizes ${\cal A}= 10 \un{\mu m^2}$
  (blue solid line),  ${\cal A}= 800 \un{\mu m^2}$ (black dotted
  line), ${\cal A}=  3\cdot 10^3 \un{\mu m^{2}}$ (green dash-dotted
  line) and ${\cal A}= 5\cdot 10^5 \un{\mu m^{2}}$ (red dashed line); In panel (b) for supra-threshold
  coupling $\kappa = 0.08 \un{mS / cm^{2}}$ and membrane sizes ${\cal
    A}= 10 \un{\mu m^2}$ (blue solid line),  ${\cal A} = 100 \un{\mu
    m^2}$ (black dashed line), ${\cal A} = 3\cdot 10^4 \un{\mu m^{2}}$
  (red dash-dotted line). For
  weak intrinsic noise the sharp peak indicates a strong 
  correlation. In contrast, in the strong noise limit the causal
  relationship is lost and the correlation function does not exhibit a
  peak. Interestingly, for the sub-threshold case a shift of the peak
  towards shorter times is observed, i.e. there is a speed up of the
  signal transmission with increasing noise level. 
}
\label{fig:correlation}
\end{figure}

In order to analyze the spike correlation between the initial and
final node, we firstly segmented the point processes $u_{0}(t)$ and
$u_{9}(t + \tau)$ in segments of width $\Delta t$. For $\Delta t$
smaller than the refractory time, there is either no spike or one
spike observable in each segment. For our analysis, we chose $\Delta t
= 1.5\un{ms}$. Secondly, we determine the number of spike coincidences
${\cal N}_{0,9}(\tau)$
between the initial node and the final node in the segments of width
$\Delta t$, i.e. the spike
coincidences between the two point processes $u_{0}(t)$ and $u_{9}(t +
\tau)$:
\begin{align}
  \label{eq:nosc}
  {\cal N}_{0,9}(\tau) = \Delta t  \, \int_{0}^{T}
  \mathrm{d} t \, f_{0}(t) f_{9}(t+ \tau)\, ,  
\intertext{where the point process $u_{i}(t)$ is approximated by}
f_{i}(t) = \int_{t}^{{t+\Delta t}} \mathrm{dt'} u_{i}(t')/\Delta t\, \text{ for }
i = 0,9 \, .
\end{align}
In Eq.~\ref{eq:nosc}, $T$ denotes the total integration time.
Note, that the above definition corresponds to a cross-correlation
measure. Thirdly, we relate this number ${\cal N}_{0,9}(\tau)$ to the number of initiated spikes at
node ``0'', i.e. ${\cal N}_{0}$, and the bin-width $\Delta t$. We
obtain the probability density $C_{0,9}(\tau)$: 
\begin{align}
  \label{eq:prop}
  C_{0,9}(\tau)= \frac{{\cal N}_{0,9}(\tau)}{\Delta t \, {\cal N}_{0}} =
  \frac{1}{{\cal N}_{0}}  \, \int_{0}^{T}
  \mathrm{d} t \, f_{0}(t) f_{9}(t+ \tau)\, .
\end{align}
Note, that due to the periodic spike initiation at the first
node, $C_{0,9}(\tau)$ is periodic, cf. Fig.~\ref{fig:correlation}. The normalization to the
total number of initiated spikes ${\cal N}_{0}$ ensures that
the integral of $C_{0,9}(\tau)$ over one period results in the transmission coefficient ${\cal R}$ defined by
Eq.~(\ref{eq:ratio}).

A sharp peak in
$C_{0,9}(\tau)$ indicates a high correlation. For
supra-threshold coupling and weak intrinsic noise, the correlation
measure $C_{0,9}(\tau)$ exhibits a sharp peak for a $\tau$-value,
which corresponds to the propagation time modulo the period of the
spiking in the initial node, cf. Fig. ~\ref{fig:correlation} (b). 
With increasing noise level, the width of the peak increases and the
height of the peak compared to the background level decreases until the
total disappearance of the peak in the background showing the lost of any causal
correlation. Interestingly, for the sub-threshold case, not only a
broadening and flattening of the peak can be observed with increasing
noise level, but also a shift of the peak towards smaller $\tau$-values showing a
speed up of the transmission, cf. Fig.~\ref{fig:correlation} (a). The
latter can be explained by the same line of reasoning as drawn by the
effect of {\itshape anticipated synchronization} in Ref.~\cite{physicaa}. 

\section{Conclusion}

We numerically studied the saltatory spike propagation along a
myelinated axon within a multi-compartmental, stochastic Hodgkin-Huxley
modeling. The channel noise affecting the dynamics of the bi-linearly
coupled nodes of Ranvier is originated in the random ion channel
gating. Upon analyzing the spike propagation in terms of transmission
reliability, i.e. the ratio
of the number of spike observed in the terminal node to the number of
spikes initiated in the first node, we found a reduction of the
transmission reliability with increasing channel noise level for
supra-threshold coupling. This is due to the noise induced propagation
failures. A further increase of the channel noise level leads to the total
loss of the spiking correlation between the first and the last node of
Ranvier of the
axonal chain. In case of sub-threshold coupling and
not so high channel noise level, the channel noise can constructively
contribute to the spike propagation and the effect of noise-assisted
spike propagation is recovered. Both the transmission reliability as
well as the propagation speed can increase with increasing channel
noise level. This observed behavior is quite in spirit of the
stochastic resonance 
phenomenon \cite{Gammaitoni,Anishenkov,hanggicpc}
 with an intrinsic noise source \cite{SchmidEPL,JungEPL} with the
 inter-nodal coupling playing the role of an nodal stimulus.  \\ 

The authors like to applaud and warmly thank Lutz Schimansky-Geier for his continuous engagement
in furthering stochastic physics within the statistical physics community worldwide
and for his many elucidative discussions with us in pursuing stochastic physics. He is still young 
and strong enough to appreciate and to contribute great science.



\begin{thebibliography}{00}
\bibitem{bb} B. B. Averbeck and D. Lee, Trends Neurosci. \textbf{27},
  225 (2004)
\bibitem{Hodgkin1952} A. L. Hodgkin and A. F. Huxley,
  J. Physiol. \textbf{117}, 500 (1952)
\bibitem{white2000} J.A. White, J.T. Rubinstein, and A.R.  Kay, Trends
  Neurosci. \textbf{23}, 131 (2000)
\bibitem{lindner} B. Lindner, J. Garc\'ia-Ojalvo, A. Neiman, and
  L. Schimansky-Geier, Phys. Rep. \textbf{392}, 321 (2004)
\bibitem{sagues} F. Sagu\'es, J. M. Sancho, and J. Garc\'ia-Ojalvo,
  Rev. Mod. Phys. \textbf{79}, 829 (2007)
\bibitem{Lecar}H. Lecar and R. Nossal, Biophys. J. \textbf{11}, 1048 (1971);
\textbf{11}, 1068 (1971) 
\bibitem{Gammaitoni} L. Gammaitoni, P. H\"anggi, P. Jung, and
  F. Marchesoni, Rev. Mod. Phys. \textbf{70}, 223 (1998) 
\bibitem{Anishenkov} V. S. Anishchenko, A. B. Neiman, F.  Moss and L. Shimansky-Geier,
Usp. Fiz. Nauk. \textbf{169}, 7 (1999) 
\bibitem{hanggicpc}
P. H\"anggi, ChemPhysChem {\bf 3}, 285 (2002) 
\bibitem{SchmidEPL} G. Schmid, I. Goychuk,  and P. H\"anggi,
    Europhys. Lett. \textbf{56}, 22 (2001)
\bibitem{JungEPL}P. Jung  and J. W. Shuai, Europhys. Lett.
  \textbf{56}, 29 (2001) 
\bibitem{schmid2004} G. Schmid, I. Goychuk, and P. H\"anggi,
Phys. Biol. \textbf{1}, 61 (2004)
\bibitem{schmid2006} G. Schmid, I. Goychuk, and P. H\"anggi,
  Phys. Biol. \textbf{3}, 248 (2006)
\bibitem{Zeng} S. Zeng, Y. Tang, and P. Jung, Phys. Rev. E
  \textbf{76}, 011905 (2007);\\
  S. Zeng and P. Jung, Phys. Rev. E
  \textbf{71}, 011910 (2005) 
\bibitem{Anna} A. Ochab-Marcinek, G. Schmid, I. Goychuk and
  P. H\"anggi, Phys. Rev. E \textbf{79}, 011904 (2009)
\bibitem{Koch}C. Koch, \textit{Biophysics of Computation, Information Processing in
Single Neurons} (Oxford University Press, New York 1999).
\bibitem{Keener}J. Keener and J. Sneyd, \textit{Mathematical Physiology} (Springer, New York 2001).
\bibitem{Fox1994} R. F. Fox and Y. N. Lu, Phys. Rev. E \textbf{49},
  3421 (1994) 
\bibitem{Tuckwell1987} H. C. Tuckwell, J. Theor. Biol. \textbf{127},
  427 (1987)
\bibitem{HanggiThomas_a} P. H{\"a}nggi and H. Thomas, Phys. Rep.
  \textbf{88}, 207 (1982); see in Sect. 6.2.
\bibitem{goychukpnas} I. Goychuk and P. H{\"a}nggi,
  Proc. Natl. Acad. Sci. \textbf{99}, 3552 (2002)
\bibitem{physicaa} G. Schmid, I. Goychuk and P. H{\"anggi}, Physica A \textbf{325}, 165 (2003)



\end{thebibliography}
\end{document}